\documentclass[twocolumn,showpacs,preprintnumbers,amsmath,amssymb,psfig]{revtex4}

\usepackage{graphicx}
\usepackage{dcolumn}
\usepackage{bm}

\newcommand{\beq}{\begin{eqnarray}}
\newcommand{\eeq}{\end{eqnarray}}
\newcommand{\beqq}{\begin{eqnarray*}}
\newcommand{\eeqq}{\end{eqnarray*}}

\begin{document}
\title
{Universal potential barrier penetration by initially confined
wavepackets} \author{Er'el Granot}
 \email{erel@yosh.ac.il}
 \affiliation{Department of Electrical and Electronics Engineering, College of Judea and Samaria, Ariel, Israel\\
}
\author{Avi Marchewka}%
 \email{marchew@post.tau.ac.il}
\affiliation{Kibbutzim College of Education\\
Ramat-Aviv, 104 Namir Road 69978 Tel-Aviv, Israel\\
\\
}
\begin{abstract}
The dynamics of an initially sharp-boundary wavepacket in the
presence of an arbitrary potential barrier are investigated. It is
shown that the penetration through the barrier is universal in the
sense that it depends only on the values of the wavefunction and
its derivatives at the boundary. The dependence on the derivatives
vanishes at long distances from the barrier, where the dynamics
are governed solely by the initial value of the wavefunction at
the boundary.

\end{abstract}
\pacs{03.65.-w, 03.65.Nk, 03.65.Xp.} \maketitle

\emph{\textbf{Introduction}} The tunneling phenomenon is a
fundamental problem in quantum mechanics. Traditionally, it is
solved and presented in most textbooks for a particle with a given
energy and momentum (see, for example \cite{Merzbacher}). Although
this treatment is usually sufficient to demonstrate the main
features of tunneling, it raises a serious difficulty, for the
uncertainty principle suggests that the particle is
instantaneously distributed over the entire space. Clearly, this
analysis implies that the particle {\it was and always will} {\it
be} at both sides of the barrier, and therefore the classical
meaning of tunneling is not clear. This problem is one of the
reasons that researchers investigated the dynamics of wavepackets,
usually Gaussians, in the process of tunneling \cite{Stamp}.
Although this treatment seems more intuitive, whenever the initial
wavepacket is analytic there is always initially a tail at the
other side of the barrier. To prevent this problem, the initial
wavepacket must be singular and to have a non-zero value {\it
only} at a finite region in space.
\\
 Clearly, tunneling is just one specific case in a much broader family of scattering problems, and
all that was said above can be generalized to transmission through any arbitrary potential.\\
During the last few years, due to the recent development in
femto-second pulses lasers \cite{Paulus,Hentschel}, optical
tweezers and atom cooling and trapping \cite{Wieman,Neuman}, an
interest in Moshinsky's shutter problem\cite{Moshinsky}was
rejuvenated\cite{Brouard,Godoy,Delgado,Granot1,Granot2}. Moshinsky
was the first to investigate the dynamics of an initially singular
wavefunction. The singularity of the initial wavefunction modeled
a fast shutter. The ability to localize particles (usually atoms)
by laser beams and then to release them instantaneously increases
the feasibility to simulate
the shutter\cite{Szriftgiser,Fort}.\\

In this paper we derive a generic formula for the Schr\"{o}dinger
dynamics of an initially singular wavefunction (which vanishes at
half space) in the presence of a finite-width potential.

It should be stressed that even if the initial wavefunction is a
smooth function (rather than singular) with a transition length
scale of $\varepsilon$, then the results presented here are valid
provided the measurements are done at distances shorter than
$t\hbar/2m\varepsilon$ (see ref.\cite{Granot1}).
\\
\emph{\textbf{The general case}} A wavefunction is confined to one
side of a potential barrier. The barrier is located around $L > 0$
and vanishes for $x < 0$ and $x \to \infty $. The initial
wavepacket, on the other hand, vanishes beyond $x > 0$ (see
Fig.1). Therefore, initially, there is no overlap between the
wavepacket and the barrier. The dynamics are governed by the
Schr\"{o}dinger equation
\begin{equation}
\label{eq1}
 - \frac{\partial ^2}{\partial x^2}\psi + V\left( {x - L} \right)\psi =
i\frac{\partial \psi }{\partial t},
\end{equation}
where $V\left( x \right)$ is the potential barrier.
Hereinafter, the units $\hbar = 1$ and $2m = 1$ are used.\\

\begin{figure}
\includegraphics[width=8cm,bbllx=110bp,bblly=540bp,bburx=440bp,bbury=740bp]{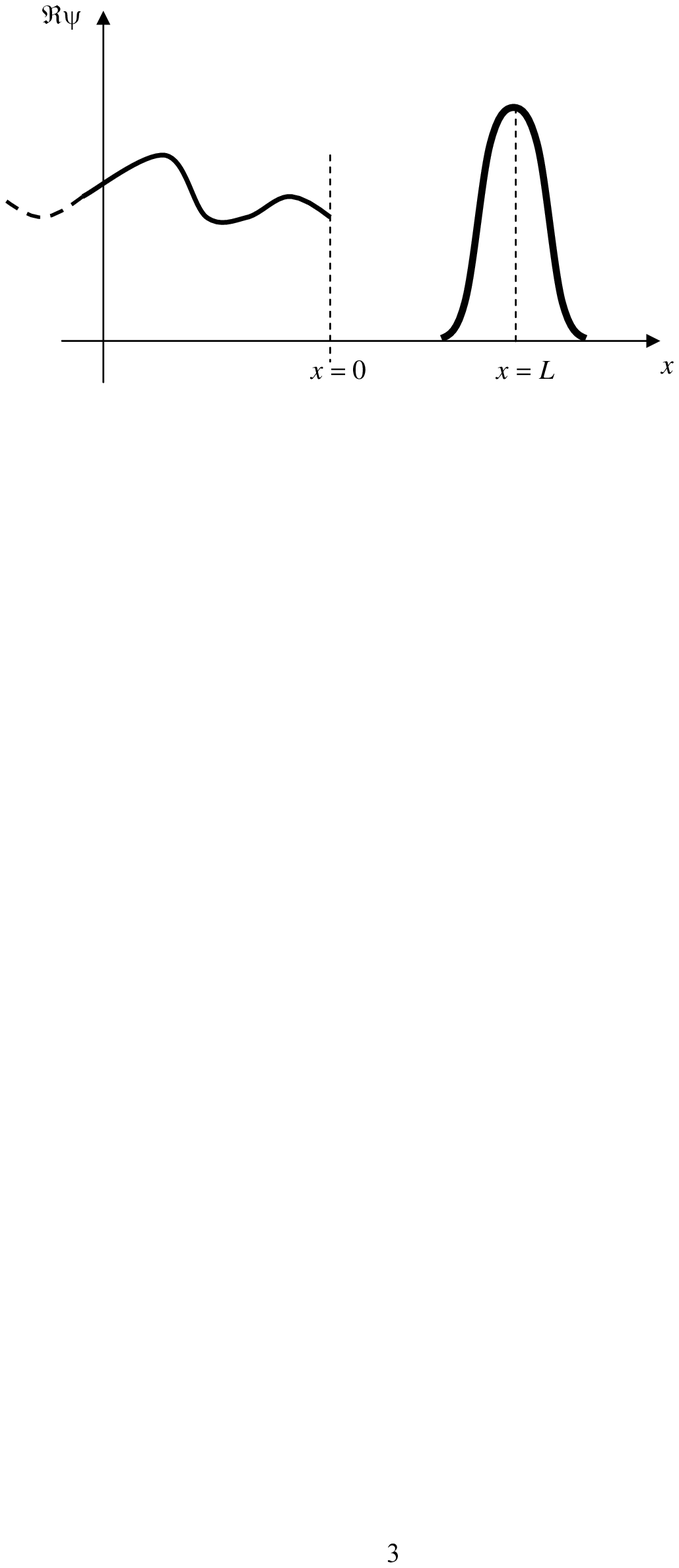}
\caption{\emph{scattering of an initially singular wavepacket over
a barrier. In the figure only the real part of the wavefunction is
plotted.}}\label{fig1}
\end{figure}

Since the initial wavefunction has a singularity at $x = 0$ it can
be written as\cite{Granot1,Granot2} $ \psi \left( {x,t = 0}
\right) = f\left( x \right)\theta \left( { - x} \right) $, and the
solution at $t > 0$ and $x > L$ is then
\begin{equation}
\label{eq3} \psi \left( {x,t} \right) = \int \frac{dkdq}{2\pi }
{F\left( q \right)\frac{iT\left( k \right)}{k - q + i0}\exp \left(
{ikx - ik^2t} \right)}.
\end{equation}
In this equation $F\left( q \right) \equiv \left( {2\pi }
\right)^{ - 1}\int {dxf\left( x \right)\exp \left( { - iqx}
\right)} $ is the Fourier transform of $f\left( x \right)$ and
$T\left( k \right)$ is the transmission coefficient of the barrier
for plane-wave with
momentum $k$.\\
The properties of the barrier and the initial wavefunction can be
separated by defining
\begin{equation}
\label{4} \varphi \left( {q,x,t} \right) \equiv \frac{1}{2\pi
}\int {dk\frac{iT\left( k \right)}{k - q + i0}\exp \left( {ikx -
ik^2t} \right)},
\end{equation}
and hence
\begin{equation}
\label{5} \psi \left( {x,t} \right) = \int {dq} \varphi \left(
{q,x,t} \right)F\left( q \right),
\end{equation}
which, by expanding $\varphi \left( {q,x,t} \right)$ in orders of
$q$, Eq. (\ref{5}) can be rewritten in terms of $f\left( x
\right)$. We thus obtain the generic solution of the problem:
\begin{equation}
\label{6} \psi \left( {x,t} \right) = \left. {\varphi \left( { -
i\frac{\partial }{\partial \xi },x,t} \right)f\left( \xi \right)}
\right|_{\xi = 0} .
\end{equation}
This generic expression should be interpreted as the infinite
expansion
\begin{eqnarray}
\label{7}
 \nonumber \psi \left( {x,t} \right) &=& \sum\limits_{n =
0}^\infty {\frac{\left( { - i} \right)^n}{n!}\frac{\partial
^n}{\partial q^n}\left. {\varphi \left( {q,x,t} \right)}
\right|_{q = 0} \frac{\partial ^n}{\partial \xi ^n}\left. {f\left(
\xi \right)} \right|_{\xi = 0} } = \\ && \varphi \left( {0,x,t}
\right)f\left( 0 \right) - i\varphi ^{\prime}\left( {0,x,t}
\right)f^{\prime}\left( 0 \right) + \cdots ,
\end{eqnarray}
where $\varphi ^{\prime}\left( {0,x,t} \right) \equiv \left.
{\partial \varphi \left( {q,x,t} \right) / \partial q} \right|_{q
= 0} $ and $f^{\prime}\left( 0 \right) \equiv
\partial \left.
{f\left( x \right) / \partial x} \right|_{x = 0} $.\\
Eqs. (\ref{6}) and (\ref{7}) are the main result of this paper. It
is evident from Eq.(\ref{7}) that the wavefunction dynamics depend
only on the initial wavepacket values at the boundary$x = 0$. It
is a universal formula in the sense that {\it any} wavefunction
$f\left( x \right)$ and {\it any} potential can be substituted in
it to find the {\it exact} dynamics beyond the potential at any $t
> 0$. In particular, when $f\left( x \right) = C$ is a constant,
then the solution is simply
\begin{equation}
\label{8} \psi \left( {x,t} \right) = C\varphi \left( {0,x,t}
\right).
\end{equation}

Eq.(\ref{7}) can also be written as the following expansion
\begin{equation}
\label{11} \psi \left( {x,t} \right) = \sum\limits_{n = 0}^\infty
{a_n f^{\left( n \right)}\left( 0 \right)},
\end{equation}
where $f^{\left( n \right)}\left( 0 \right) \equiv \left.
{\partial ^nf\left( \xi \right) / \partial \xi ^n} \right|_{\xi =
0} $ are the initial function's derivatives and the coefficients
are
\begin{equation}
\label{12} a_n \equiv \frac{i^{n + 1}}{2\pi n!}\int
{dk\frac{T\left( k \right)}{\left( {k + i0} \right)^{n + 1}}\exp
\left( {ikx - ik^2t} \right)} .
\end{equation}
These coefficients can be written also as an expansion by
expanding $T\left( k \right) / \left( {k + i0} \right)^{n + 1}$
around the momentum $x / 2t$ (the dominant term in the stationary
phase approximation in long times and arbitrary$x)$, i.e.,
\begin{eqnarray}
\label{13} \frac{T\left( k \right)}{\left( {k + i0} \right)^{n +
1}} =\sum\limits_{m = 0}^\infty {\frac{\left( {k - x / 2t}
\right)^m}{m!}\left. {\frac{\partial ^m\left[ {T\left( q \right) /
q^{n + 1}} \right]}{\partial q^m}} \right|_{q = x / 2t} }.
\end{eqnarray}
By substituting Eq.(\ref{13}) in Eq.(\ref{12}) and Eq.(\ref{11})
one obtains
\begin{equation}
\label{14} \psi \left( {x,t} \right) = \frac{1}{2\pi }\exp \left(
{i\frac{x^2}{4t}} \right)\sum\limits_{n = 0}^\infty {b_n f^{\left(
n \right)}\left( 0 \right)},
\end{equation}
where the coefficients are
\begin{equation}
\label{15} b_n \equiv \frac{i^{n + 1}}{n!}\sum\limits_{m =
0}^\infty {\frac{s_{n,2m} }{\left( {2m} \right)!}\Gamma \left( {m
+ \textstyle{1 \over 2}} \right)\left( {it} \right)^{ - \left( {2m
+ 1} \right) / 2}} ,
\end{equation}
and $s_{n,2m} \equiv \frac{\partial ^{2m}}{\partial k^{2m}}\left[
{T\left( k \right)k^{ - n - 1}} \right]_{k = x / 2t} $. This
result also suggests that at large distances $x^2 / t \to \infty $
\begin{equation}
\label{16} \psi \left( {x,t} \right)\sim T\left( {\frac{x}{2t}}
\right)\sqrt {\frac{it}{\pi }} \frac{e^{ix^2 / 4t}}{x}f\left( 0
\right),
\end{equation}
and the particles density decays like $\rho = \left| {\psi \left(
{x,t} \right)} \right|^2\sim \left| {T\left( {x / 2t}
\right)f\left( 0 \right)} \right|^2tx^{ - 2} / \pi $, which agrees
with the free results\cite{Granot1} for $T = 1$. Moreover, for
every finite potential $T\left( {k \to \infty } \right) \to 1$,
which means that at very short times (with respect to the
distance, i.e., $t / x^2 < < 1$, although it was derived for large
$t)$ Eq. (\ref{16}) is reduced to the free propagation for every
potential, which is consistent with \cite{Granot2}.

\begin{figure}
\includegraphics[width=8cm,bbllx=75bp,bblly=540bp,bburx=444bp,bbury=765bp]{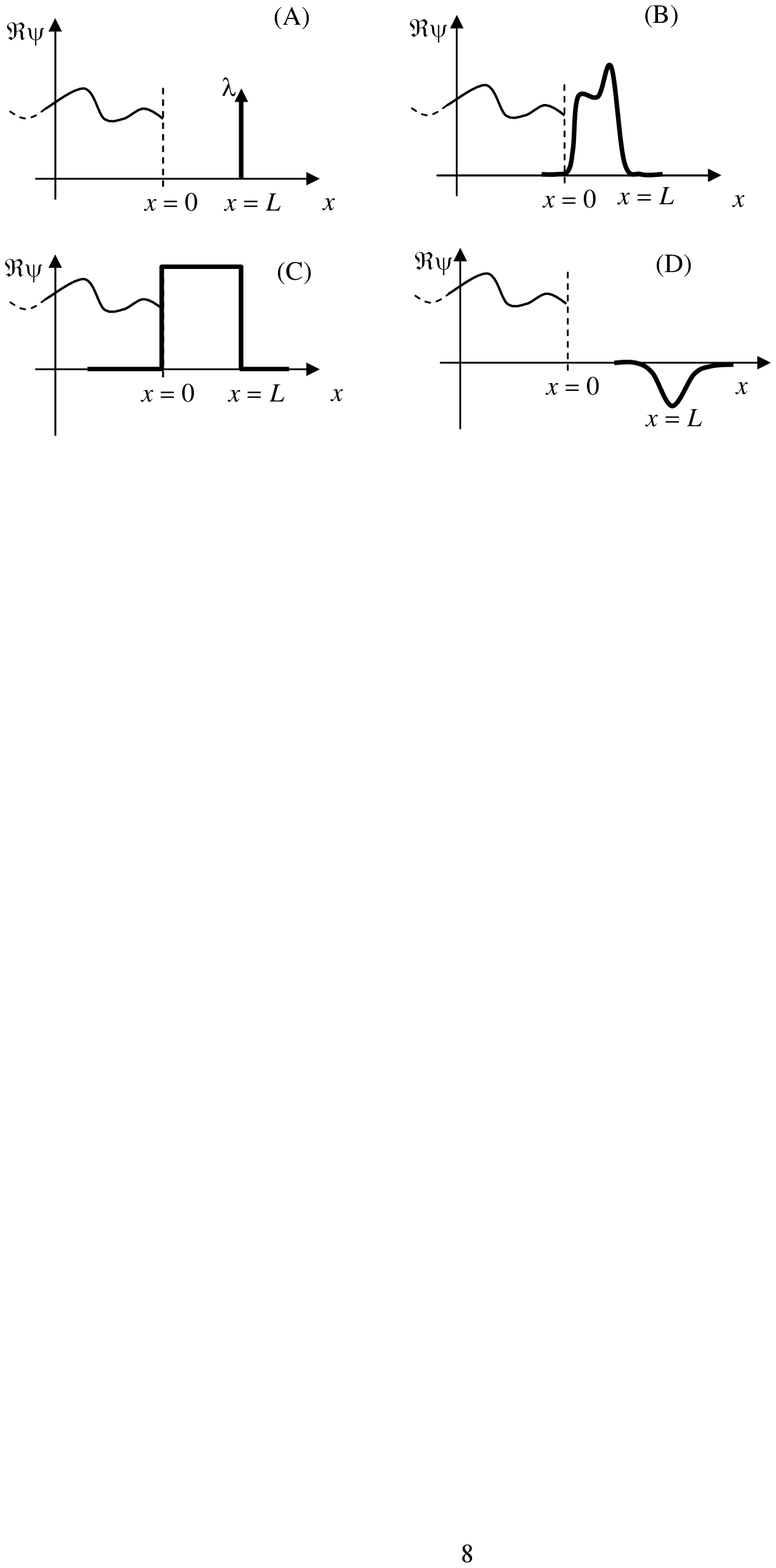}
\caption{\emph{An illustration of the four scenarios under study.
A) delta function potential B) Arbitrary smooth opaque barrier C)
Rectangular barrier and D) Reflectionless potential
well}}\label{fig2}
\end{figure}

\emph{Specific examples and applications} In this section we will
demonstrate these derivations in a few scenarios for different
potentials.
\\
\emph{\textbf{The delta function potential -- an exact solution.}}
One of the simplest scenarios of a potential barrier penetration
is the delta-function potential (see Fig.2A). This problem was
discussed in detail for an initially singular wave function in
\cite{Granot2}, and is presented here only to illustrate the
validity of the above derivations. In this case the
Schr\"{o}dinger equation reads: $
 - \frac{\partial ^2}{\partial x^2}\psi + \lambda \delta \left( {x - L}
\right)\psi = i\frac{\partial \psi }{\partial t} $. The
transmission coefficient in this case is well known $ T\left( k
\right) = k/(k - i\lambda / 2)$, in which case by Eqs.(\ref{4})
and (\ref{6}) the solution beyond the barrier is
\begin{widetext}
\begin{eqnarray} \label{eq20}
 \psi \left( {x > L,t} \right) =
 \frac{e^{ix^2 / 4t}}{2}\left\{ {\frac{\partial / \partial \xi }{\partial /
\partial \xi + \lambda / 2}w\left[ {\sqrt {it} \left( {\frac{x}{2t} +
i\frac{\partial }{\partial \xi }} \right)} \right] - \frac{\lambda
/ 2}{\partial / \partial \xi + \lambda / 2}w\left[ {\sqrt {it}
\left( {\frac{y}{2t} - i\frac{\lambda }{2}} \right)} \right]}
\right\}\left. {f\left( \xi \right)} \right|_{\xi = 0}.
 \end{eqnarray}
\end{widetext}
This expression obviously agrees with \cite{Granot2}, and
illustrates the fact that the dynamics are fully described by the
values of the
initial wavefunction at the singular point.\\
At short times, i.e., $t < < x^2$, it can be expanded (see
\cite{Granot2})
\begin{eqnarray}
\label{eq21} &&\psi \left( {x,t} \right) \cong \\\nonumber
&&\sqrt{\frac{it}{\pi }} \frac{e^{ix^2 / 4t}}{x}\left[ {\left( {1
+ \frac{i\lambda }{x}t} \right)f\left( 0 \right) -
i\frac{2t}{x}f^{(1)}\left( 0 \right) + O\left( {x^{ - 2}} \right)}
\right],
\end{eqnarray}
This expression is, of course, consistent with Eqs. (\ref{11}) and (\ref{12}).\\
A special case occurs when $\lambda = 2f^{(1)}\left( 0 \right) /
f\left( 0 \right)$. In this case the leading terms have no
dependence on the barrier:
\begin{equation}
\label{eq22} \psi \left( {x,t} \right) \cong \sqrt {\frac{it}{\pi
}} \frac{e^{ix^2 / 4t}}{x}\left[ {f\left( 0 \right) + O\left( {x^{
- 2}} \right)} \right].
\end{equation}
That is, in the leading terms, the influence of the barrier was
eliminated by changing the derivative of the initial wavefunction
only at the singularity point for $\sqrt t / x < < 1$. In general,
each one of the terms in the expansion (\ref{eq21}) can be
eliminated (except for the first one) by tailoring the derivatives
of the wavefunction at the singular point.
\\
\emph{\textbf{An arbitrary smooth opaque barrier.}} In the case of
an opaque barrier with smooth edges (see Fig.2 B) one can use the
WKB approximation to evaluate the transmission
coefficient of the barrier.\\
If the potential function is $V\left( x \right)$ then
\cite{Merzbacher} $ T\left( k \right) = \frac{2}{\left( {2\theta +
1 / 2\theta } \right)} $, where $\theta \left( k \right) \equiv
\exp ( {\int_0^L {d\eta [V\left( \eta \right) - k^2]^{1/2} } } )$.
Therefore, from Eq.(\ref{14})
\begin{equation}
\label{eq24} \psi \left( {x,t} \right)\sim \sqrt {\frac{it}{\pi }}
\frac{f\left( 0 \right)}{x}\frac{2\exp \left( {ix^2 / 4t}
\right)}{\left[ {2\theta \left( {x / 2t} \right) + 1 / 2\theta
\left( {x / 2t} \right)} \right]}.
\end{equation}
Eq. (\ref{eq24}) is valid, of course, only when the turning points
exist, i.e., when $\left( {x / 2t} \right)^2 < \max \left\{
{V\left( x
\right)} \right\}$.\\
For very opaque barrier, i.e., $\int_0^L {d\eta [V\left( \eta
\right) - \left( {x / 2t} \right)^2]^{1/2} } \gg 1$,
\begin{equation}\nonumber
\psi \left( {x,t} \right)\sim \sqrt {\frac{it}{\pi }}
\frac{f\left( 0 \right)e^{ix^2/4t}}{x}\theta \left(
\frac{x}{2t}\right).
\end{equation} \emph{\textbf{The rectangular
potential barrier.}} In the case of the rectangular potential
barrier (see Fig.2C), with potential height and length $V_0 $ and
$L$ respectively, the transmission coefficient is\cite{Merzbacher}

\begin{equation} T\left( k \right) = \frac{\exp \left( { - 2ikL}
\right)}{\cosh \left( {2\kappa L} \right) + i\left( {\varepsilon /
2} \right)\sinh \left( {2\kappa L} \right)} \end{equation}, where
$\kappa \equiv \sqrt {V_0 - k^2} $ and $\varepsilon \equiv \kappa
/ k - k / \kappa $.

At short times, the particles that penetrate the barrier are
extremely energetic and therefore they pass the barrier almost
unaffected by it.\\
In this case $\kappa \cong ix / 2t$, $\varepsilon \cong i\left[ {1
+ 2\left( {V_0 t^{2} / x^2} \right)^2} \right]$ and
\begin{eqnarray}\nonumber
\label {eq27} &&T\left( {\frac{x}{2t}} \right) \cong \left[1 -
2i\left( {V_0 t^2 / x^2} \right)^2\sin \left( {xL / t} \right)\exp
\left( {ixL / t} \right) \right]^{-1},
\end{eqnarray}
 which
yields by Eq.(\ref{16})
\begin{eqnarray}
\label{eq27} &&\psi \left( {x,t} \right) \to \frac{e^{ix^2 /
4t}f\left( 0 \right) \sqrt{it/\pi}/x}{1 - 2i\left( {V_0 t^2 / x^2}
\right)^2\sin \left( {xL / t} \right)e^{ixL / t}},
\end{eqnarray}

with weak resonances at $xL / t = m\pi $, (where $m$ is an
integer).
\\
\emph{\textbf{A Reflectionless potential.}} One of the peculiar
potential examples is the reflectionless
potential\cite{Landau,Tong}. For this potential (see Fig. 2D) the
absolute value of the transmission coefficient is always $\left|
{T\left( k \right)} \right| = 1$. Therefore, the reflection
coefficient is zero for every incoming plane wave; however, the
transmission coefficient suffers from dispersion, which deforms
the initial wavepacket.\\
It is well known that the potential $ V\left( x \right) = -
2a^2\mathrm{sech}^2\left( {ax} \right) $, whose width goes like
$\sim a^{ - 1}$ and its depth like $a^2$,
belongs to a family of reflectionless potentials \cite{Landau,Kiriushcheva,Tong}.\\
This potential, as opposed to the previous three, is not
completely localized in space. That is, the initial wavefunction
"feels" the potential at any distance. Therefore, one should place
it at a large distance from the edge of the initial wavefunction.\\
Its Schr\"{o}dinger equation is then
\begin{equation}
\label{eq30}
 - \frac{\partial ^2}{\partial x^2}\psi - 2a^2\mathrm{sech}^2\left[ {a\left( {x -
L} \right)} \right]\psi = i\frac{\partial \psi }{\partial t},
\end{equation}
where $L > > a^{ - 1}$ is the new location of the potential well.
This equation has an exact reflectionless solution (which
corresponds to the incoming plane wave $\psi \left( {x \to -
\infty } \right)\sim \exp \left( {ikx} \right) + R\left( k
\right)\exp \left( { - ikx} \right)$ )
\begin{equation}
\label{eq31}
\psi \left( x \right) = \left[ {k / a + i\tanh \left( {ax} \right)}
\right]\left( {k / a - i} \right)^{ - 1}\exp \left( {ikx} \right).
\end{equation}

For $x \to \infty $ Eq.(\ref{eq31}) can be written $\psi \left( x
\right)\sim T\left( k \right)\exp \left( {ikx} \right)$ with the
simple transmission coefficient\\
$ T\left( k \right) = \frac{k + ia}{k - ia}. $ According to
Eqs.(\ref{6}),

$
\varphi \left( {q,x > L,t} \right) = \frac{1}{2\pi }\int\limits_{}
^{}{dk} \frac{i}{k - q + i0}\frac{k + ia}{k - ia}\exp \left( {ikx
- ik^2t} \right),
$
which can easily be written as \cite{Granot2}
\begin{eqnarray}
\label{eq33} \varphi \left( {q,x,t} \right)& =& \varphi _{free}
\left( {q,x,t} \right) - \\\nonumber
 &&\frac{2a}{q - ia}\left[
{\varphi _{free} \left( {q,x,t} \right) - \varphi _{free} \left(
{a,x,t} \right)} \right],
\end{eqnarray}
where $\varphi _{free} \left( {q,x,t} \right)$ is the free
(without the barrier) solution $ \varphi _{free} \left( {q,x,t}
\right) = \frac{1}{2}w\left[ {\sqrt {it} \left( {\frac{x}{2t} - q}
\right)} \right]\exp \left( {ix^2 / 2t} \right) $. The solution
according to Eq.(\ref{6}) is finally
\begin{widetext}
\begin{eqnarray}
\label{eq36}
\psi \left( {x,t} \right) = \left[ {\varphi _{free}
\left( { - i\frac{\partial }{\partial \xi },x,t} \right)
-\frac{2ai}{\partial /
\partial \xi + a}\left[ {\varphi _{free} \left( { - i\frac{\partial
}{\partial \xi },x,t} \right) - \varphi _{free} \left( {a,x,t}
\right)} \right]} \right]\left. {f\left( \xi \right)} \right|_{\xi
= 0}
\end{eqnarray}.
\end{widetext}

In a particular case, where there is a discontinuity only in the function,
and not in its derivatives, i.e., if the initial wavefunction looks like
$\psi \left( {x,t = 0} \right) = f\left( 0 \right)\theta \left( { - x}
\right)$, then
\begin{eqnarray}
\label {eq37} &&\psi \left( {x,t} \right) = \frac{1}{2}f\left( 0
\right)\exp\left(i\frac{x^2}{2t}\right)\times \\\nonumber
&&\left\{ {w\left( {\frac{x}{2}\sqrt {\frac{i}{t}} } \right)
-2i\left[ {w\left( {\frac{x}{2}\sqrt {\frac{i}{t}} } \right) -
w\left( {\sqrt {it} \left( {\frac{x}{2t} - a} \right)} \right)}
\right]} \right\}.
 \end{eqnarray}
 Again, at short times
$t < < x / a$ the free solution is retrieved (the two expressions
on the right cancel each other).
\\
\emph{\textbf{Conclusions and summary}}

We have presented a generic formula, which describes the solution
of the temporal propagation through an {\it arbitrary} potential
of an initially singular wavefunction. It was shown that when the
initial wavefunction vanishes at the entire half space, i.e.,
$\psi \left( {x,t = 0} \right) = f\left( x \right)\theta \left( {
- x} \right)$ the wavefunction for every $t$ is $\psi \left( {x,t}
\right) = \left. {\varphi \left( { - i\frac{\partial }{\partial
\xi },x,t} \right)f\left( \xi \right)} \right|_{\xi = 0} ,$where
$\varphi \left( {q,x,t} \right) \equiv \frac{1}{2\pi
}\int\limits_{} ^{} {dk\frac{i}{k - q + i0}T\left( k \right)\exp
\left( {ikx - ik^2t} \right)} $ is a function, which depends only
on the transmission coefficient of the barrier $T\left( k
\right)$. This expression was also generalized to an initially
compact support wavefunction.\\

Particularly, it is shown that at very large distances the
solution can be approximated by the generic expression for any
initial function $\psi \left( {x,t = 0} \right) = f\left( x
\right)\theta \left( { - x} \right)$ the general solution is $
\psi \left( {x,t} \right)\sim T\left( {\frac{x}{2t}} \right)\sqrt
{\frac{it}{\pi }} \frac{e^{ix^2 / 4t}}{x}f\left( 0 \right). $

The authors are indebted to Miriam Schler for her networking
efforts.

\end{document}